# Differential deposition applied to x-ray mirror substrates

Patrice Bras*[a], Sylvain Labouré[a], Amparo Vivo[a], François Perrin[a], Christian Morawe[a]
[a]European Synchrotron Radiation Facility, 71, avenue des Martyrs, 38043 Grenoble, France

## ABSTRACT

The process of differential deposition is currently applied at the ESRF in order to correct figure errors of x-ray optics substrates, prior to multilayer deposition. The substrate is moved at a controlled speed in front of a sputtering source to precisely control the deposition profile. This work will describe the concept of differential deposition at the ESRF as well as recent results of its implementation to correct a real mirror substrate surface. Finally, initial studies using a synchrotron beamline characterization technique based on x-ray total reflection are presented.

## 1. INTRODUCTION

Novel generation low-emittance x-ray sources require optics exhibiting state-of-the-art surface figure. In particular, height errors well below 1 nanometer are becoming standard in order to preserve the integrity of the incident x-ray beam. Standard chemical-mechanical polishing techniques can be optimized to reach height-errors in the nanometer range. Going further usually implies the use of a combination of advanced techniques such as ion-beam figuring (IBF) [1], elastic-emission machining (EEM) [2], or differential deposition (DD) [3]. This paper presents how differential deposition is implemented at the ESRF multilayer laboratory as well as some recent results. A characterization technique based on x-ray total reflection at the ESRF bending magnet beamline BM05 is also introduced as a way to assess the effectiveness of the surface correction process.

## 2. METHODOLOGY

### 2.1 Differential deposition

DD relies on moving a substrate in front of particle flux with a given speed profile. A mask with a well-defined aperture is placed between the particle source and the substrate to restrict the particle flux lateral extent. In the present application, sputtering-based DD is used to correct the surface figure of x-ray mirrors. Figure 1 shows the procedure used to calculate the required velocity profile, starting from surface height error measurements of the targeted substrate.

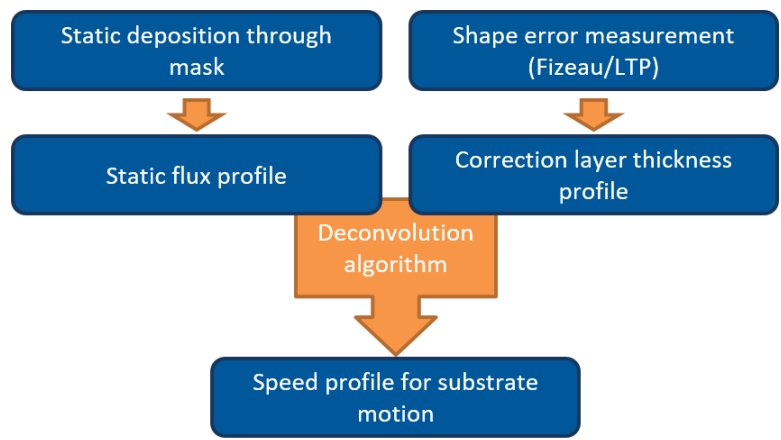

**Figure 1. Substrate velocity profile calculation process for differential deposition**

On the one hand, the static particle flux obtained through the used apertures is first determined. The approach followed relies on two different lateral apertures, 24 mm and 2 mm, to correct surface features of large and relatively smaller lateral periods, respectively.

On the other hand, the surface height error profile of the device of interest is determined using optical metrology equipment, mainly the long-trace profiler (LTP) developed at the ESRF [4]. The corrective layer thickness profile is then calculated. The static particle flux and the thickness profile are used as inputs in a deconvolution algorithm to finally obtain the required velocity profile to be applied to the substrate during DD. A more detailed description is available in [5].

The material used for DD is $WSi_2$ as it exhibits suitable properties in thin film form, namely low surface roughness and limited stress [6].

The standard procedure implemented here relies on a 2-step correction. The surface figure is initially characterized with visible light metrology. A first correction is performed using the 24 mm aperture. After another surface height error measurement, a second correction is performed, this time using the 2 mm aperture.

## 2.2 Total reflection imaging

Keeping in mind that the corrected substrates are to be used on beamlines eventually, beamline-based characterization techniques can give important insights. Figure 2 presents the proposed setup, as a way to assess the impact of successive correction iterations.

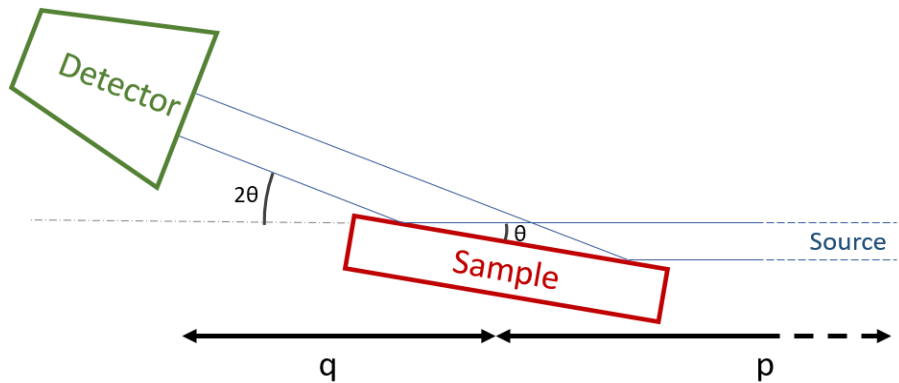

**Figure 2. Total reflection imaging setup at BM05**

This measurement consists of total reflection imaging at a photon energy of 11 keV, in theta-2theta configuration for shallow angles. One image is typically taken every 0.01° for incidence angles θ between 0 and 0.8°. The 2-D detector used in this experiment (PCO Edge Gold 4.2) has a pixel size of 0.65 x 0.65 µm² and a field of view of 1.33 x 1.33 mm². The source-mirror distance, denoted p in Figure 2, is 60 metres. The mirror-detector distance is q = 280 mm. The incident angle should exceed θ = 0.2° to avoid the impact of the direct beam on the detector.

# 3. RESULTS

## 3.1 Differential deposition

Static deposition profiles obtained with the 24 mm and 2 mm apertures are shown in Figure 3.

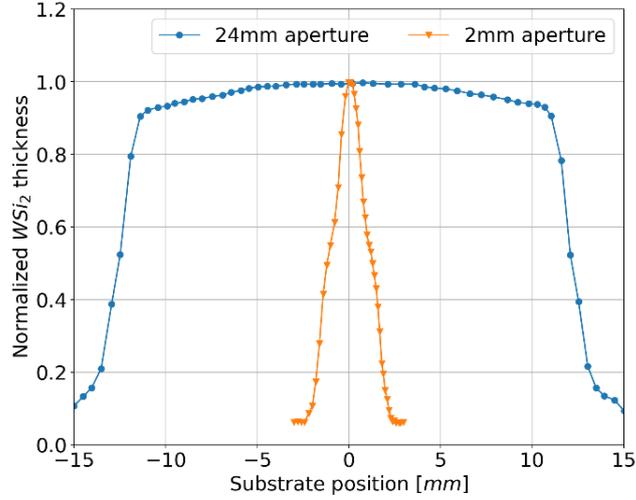

**Figure 3.  Normalized WSi$_2$ static deposition profiles with 24 mm (blue), and 2 mm (orange) apertures**

Both profiles exhibit lateral shoulders which correspond to images of the target erosion lines through the mask. The full width at half maximum are 25 mm and 2.6 mm respectively.

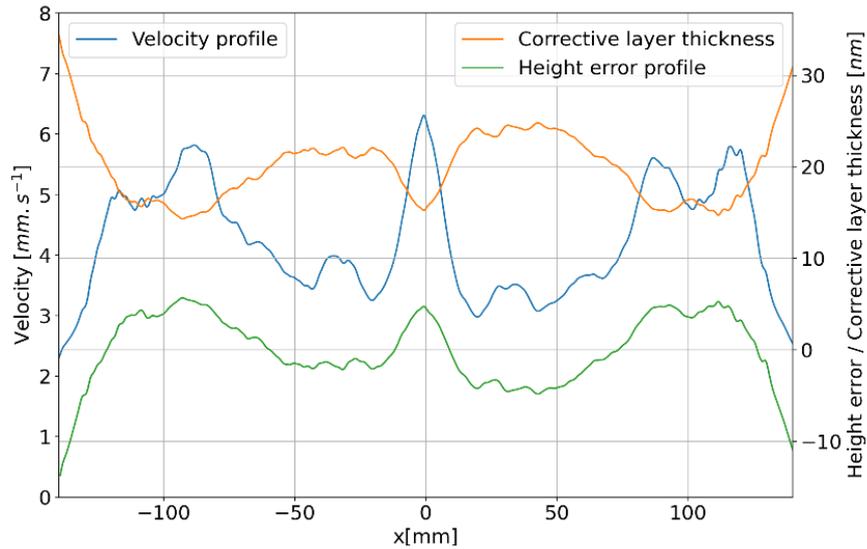

**Figure 4. Height error profile (green), calculated corrective layer thickness (orange) and corresponding substrate velocity profile (blue) during differential deposition.**

Figure 4 shows the initial height errors on the substrate measured with the LTP (green), the corresponding corrective layer thickness profile (orange) and the velocity profile to be applied during differential deposition (blue), calculated using the procedure shown in Figure 1. As one could expect intuitively, variations of the speed profile follow the same trend as the height errors, as more material, i.e. low motion velocity, is needed to fill the valleys. Conversely, the substrate velocity is increased where the height errors are more positive.

A summary of surface figure characterization after successive corrections is shown in Figure 5.

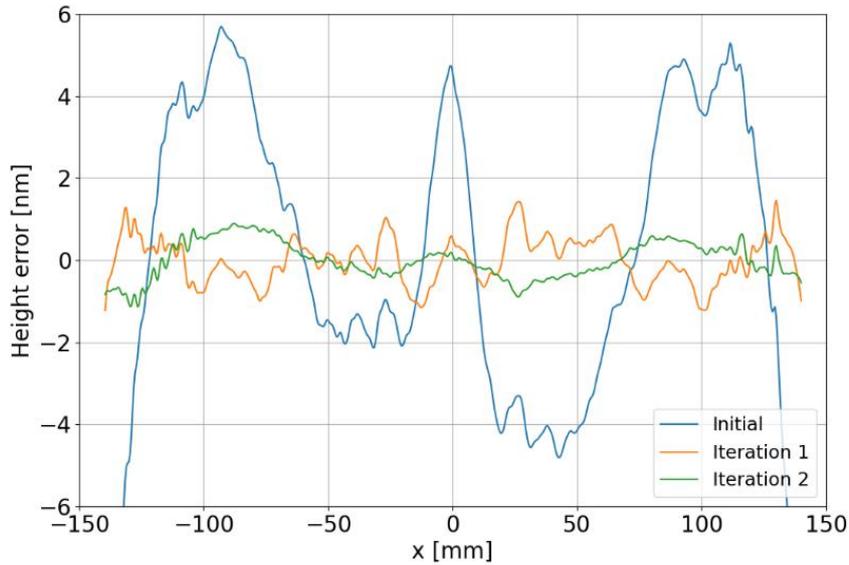

**Figure 5. Height errors on the initial surface (blue), after the 1$^{st}$ (orange) and the 2$^{nd}$ (green) correction by differential deposition.**

As expected, the first correction reduces long-wavelength contributions to the height errors while the second iteration is more effective to correct low-wavelength features. Table 1 summarizes height error values at each step of the correction process.

**Table 1. Height errors measured on the initial surface, after the 1st and the 2nd correction.**

| Correction iteration | Height errors [nm] | |
|---|---|---|
|  | RMS | PV |
| Initial | 3.87 | 20.1 |
| After 1$^{st}$ iteration | 0.57 | 3.09 |
| After 2$^{nd}$ iteration | 0.46 | 2.03 |

Height errors are significantly reduced after two correction iterations by differential deposition, from 3.87 nm down to 0.46 nm (RMS) and from 20.1 nm down to 2.03 nm (PV).

### 3.2 Total reflection imaging

Part of the results obtained at ESRF BM05 beamline in total reflection configuration are shown in Figure 6. Images taken at 0.2° incidence angle are presented for the initial surface (a), after the first correction (b) and after the second correction (c). It is important to notice that the average beam intensity is different for the three images as measurements were carried out on three different dates with different beam configurations. Marks on the beamline optics contaminate the images. Horizontal stripes are visible in all three images, especially towards the bottom and the top of the images. The latter are very likely caused by diffraction effects from the mirror edges.

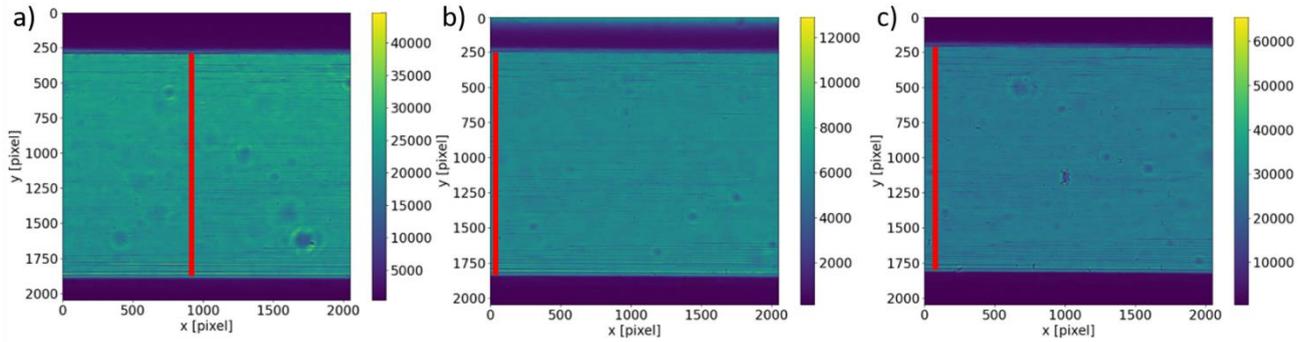

**Figure 6.** Images obtained in total reflection at an incidence angle θ of 0.2° on the initial surface (a), after the 1$^{st}$ correction (b) and the 2$^{nd}$ correction (c). The red line represent the same longitudinal profile, shown in Figure 7.

Based on careful image analysis, the same longitudinal profile, materialized by the red line in Figure 6, was extracted, normalized and compared (Figure 7).

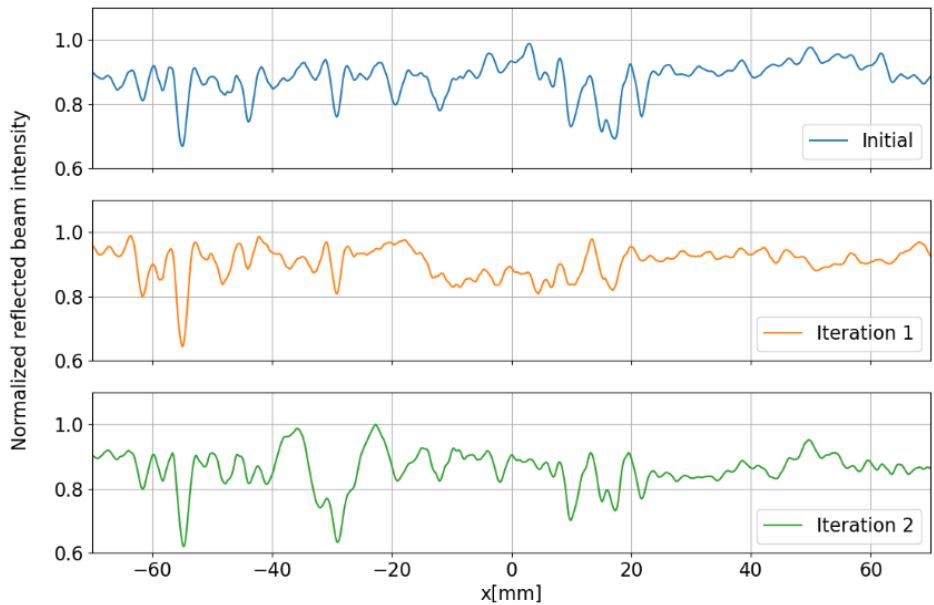

**Figure 7.** Longitudinal reflected intensity profile extracted from images shown in Figure 6.

Focusing on the location of peaks and valleys, it appears clear that the same longitudinal profile is presented in each case. The surface height errors of x-ray optics are supposed to have a detrimental impact on the reflective behavior [7]. Improving a given surface figure is expected to enhance and smooth out its reflective behavior. Here, no significant improvement is visible in the normalized reflected beam intensity after correction. Detrimental contributions from different forms of surface contaminations, observed in visible light metrology (not shown) could explain these findings. Better quality mirrors and further studies will be required to validate this characterization method. Simulation techniques may be applied to reconstruct the surface height errors from a given intensity profile.

## 4. CONCLUSION

Differential deposition was successfully implemented as a mean to correct the surface figure errors of x-ray mirrors. Using a 2-step correction process, both high and low wavelength contributions to surface figure can be mitigated. Height errors are successfully reduced down to 0.46 nm RMS. A beamline characterization technique based on total reflection imaging of the mirror is introduced as a potential method to visualize the impact of successive corrections. Further investigations are needed, on higher quality substrates, to confirm the relevance of this approach.

## ACKNOWLEDGEMENTS

The authors would like to thank Thu Nhi Tran Caliste and Luca Capasso from ESRF BM05 beamline for assistance during experiment set-up and data acquisition.